\documentclass[a4paper,12pt]{article}
\usepackage[english]{babel}
\usepackage[utf8]{inputenc}
\usepackage{amsmath, amsfonts, mathrsfs, latexsym, multicol} 
\usepackage{pgf,epsfig,graphicx,pdfsync,url}
\usepackage{units}
\usepackage{enumerate}
\newcommand{\Hs}{\mathfrak{H}}
\newcommand{\Hv}{{\mathsf{H}}}
\newcommand{\hs}{{\mathfrak{h}}}
\newcommand{\hv}{{\mathsf{h}}}
\newcommand{\Sv}{{\mathsf{S}}}
\newcommand{\sv}{{\mathsf{s}}}
\newcommand{\sper}{{\mathsf{r}}}
\newcommand{\tsper}{\tilde{\mathsf{r}}}

\newcommand{\sigmav}{{\bm{\sigma}}}
\usepackage{bm}
\newcommand{\tr}{\mathop{\mathrm{tr}}}
\providecommand{\ket}[1]{\left \vert #1 \right \rangle}
\providecommand{\bra}[1]{\left \langle #1 \right \vert}
\providecommand{\expc}[1]{\left \langle #1 \right \rangle}
\title{Resonance phenomena in the interaction of a many-photon wave packet and a qubit}
\author{Omri Gat${}^*$, Max Lein${}^{**}$ \& Stefan Teufel${}^{\dagger}$}

\date{\today}

\begin{document}

\maketitle
\vspace{-9mm}
\begin{center}
	$^{\ast}$ Hebrew University of Jerusalem, Racah institute of Physics, \linebreak
	Jerusalem 91904, Israel. \linebreak
	{\footnotesize \texttt{omrigat@cc.huji.ac.il}}
	\\[3mm]
	$^{\ast\ast}$ Kyushu University, Faculty of Mathematics, \linebreak
	744 Motooka, Nishi-ku, Fukuoka-shi, Fukuoka-ken, Japan. \linebreak
	 {\footnotesize \texttt{lein@ma.tum.de}}
	\\[3mm]
	$^{\dagger}$ Eberhard Karls Universit\"at T\"ubingen, Mathematisches Institut \linebreak
	Auf der Morgenstelle 10, 72076 T\"ubingen, Germany. \linebreak
	{\footnotesize \texttt{stefan.teufel@uni-tuebingen.de}}
\end{center}
\begin{abstract}
We demonstrate that a highly excited quantum electromagnetic mode strongly interacting with a single qubit exhibits several distinct resonances in addition to the Bloch-Siegert resonance condition that arises in the interaction with classical radiation. The resonance phenomena are associated with non-classical effects: Collapse of Rabi oscillations, splitting of the photon wave packet, and field-qubit entanglement. The analysis is based on a semiclassical phase-space flow of 2-by-2 matrices and becomes exact when the photon number tends to infinity. The flow equations are solved perturbatively and numerically, showing that the resonant detuning of the field and qubit frequencies lie on two branches in the leading order, that further split into distinct curves in higher orders.
\end{abstract}

\section{Introduction}
Physical systems where two-level systems dipole-interact with the quantum electromagnetic field are ubiquitous. Once the strong-interaction condition is attained, where the qubit-field interaction is appreciably stronger than dissipative environment coupling, these systems facilitate the coherent manipulation of qubits. A pioneering setup is cavity quantum electrodynamics (QED) \cite{Raimond:2001ts}, where atoms interact with electromagnetic modes of a high-Q cavity. An analogous process has been realized in superconducting circuits, where qubits based on Josephson junctions interact with the electromagnetic field in wave guides \cite{Wallraff:2004dy,Katz:2006ky}.

The atom-field interaction in cavity QED is often very weak, so that efficient interaction can only be achieved by making it \emph{resonant}, in the sense that the frequency of the electromagnetic mode is nearly equal to the energy spacing of two atomic levels (divided by Planck's constant). Under these conditions the amplitude of atomic levels beside the two resonant ones is negligible, and the atom evolves as a qubit; the same conditions also imply that the effect of the nonresonant terms in the field-atom interaction Hamiltonian is very weak, motivating the `rotating wave approximation' (RWA), wherein these nonresonant terms are dropped from the Hamiltonian \cite{Jaynes:1963zz}. 

Nevertheless as the interaction becomes stronger, the effects of the non-resonant terms becomes noticeable and leads to qualitative changes in the dynamics. These effects were first studied by Bloch and Siegert in the context of nuclear magnetic resonance where the spin interacts with classical electromagnetic radiation \cite{Bloch:1940wy}. They showed that the resonance curve of the Rabi oscillations of the spin polarization as a function of the radiation frequency shifts so that its center is different from the qubit natural frequency; the qubit-field resonant detuning is proportional to the square of the interaction strength.

By the correspondence principle, when the quantum electromagnetic mode is prepared in a highly excited coherent state the system dynamics should be close to that of a qubit in a classical field. This is indeed the case for short times, where the field evolution is approximately classical and the qubit Rabi oscillates, but for longer times manifestly non-classical behavior takes place, including collapse-revival of the Rabi oscillations \cite{Eberly:1980tj}, and splitting of the electromagnetic field wave packet to create `Schr\"odinger cat' states \cite{GeaBanacloche:1990tz,GeaBanacloche:1991tr}. Nevertheless these effects can be understood and described with high accuracy using \textit{semi}classical theories based on flow in generalized phase spaces \cite{Gat:1101871,Leshem:2011ji}.

Most of the studies of highly excited field cavity QED were carried out in the RWA on resonance. When the qubit-field interaction is not very small, the RWA is no longer valid and the resonance shifts, and, as shown below, splits into several distinct resonances. There is a large literature on the physics of the full Rabi model, see for example \cite{Schweber:1967wp,Graham:1984wh,Irish:2005wx,Irish:2007wu,Hausinger:2008fg,Liu:2009vm,Hausinger:2010eb,Casanova:2010wy,Ashhab:2010ta,Braak:2011hc, Beaudoin:2011ip,Crespi:2012im}, but fewer studies have been devoted so far to the interaction with highly excited initial wave packets \cite{Zaheer:1988wb,Muller:1991ut,Finney:1994wr}. These works used semiclassical phase space flow to analyze the collapse and revival of Rabi oscillations, wave packet splitting and entanglement, but the semiclassical methods available at the time were not sufficient to develop a systematic approximation.

Here we present a general semiclassical framework for the dynamics of a canonical degree of freedom coupled to a two-level system based on the Wigner-Weyl phase space representation with matrix-valued symbols \cite{Littlejohn:1991uq}. Our theory is based on the propagation of matrix-valued phase-space symbols of smooth observables in the Heisenberg picture \cite{Gat:1101871,Littlejohn:1991uq,Teufel:2003vf,Sundaram:1999wh,Panati:2003ul}. The symbols are expanded in the semiclassical small parameter $\varepsilon$, inversely proportional to the mean photon number, and a hierarchy of partial differential equations is obtained from the Heisenberg equations of motion. We solve the equations to subleading order in $\varepsilon$, where the phase-space derivatives are first-order and the equations therefore describe a flow in phase space. The leading-order equations describe the classical dynamics of the electromagnetic field mode and the Rabi oscillations of the qubit in a classical electromagnetic field, while the subleading flow equations describe the splitting of the wave packet.

The theory developed here is different from that of the well-studied semiclassical mechanics of many-component wave function \cite{Littlejohn:1991uq,Sundaram:1999wh,Panati:2003ul} in that the leading term in the Hamiltonian is scalar, that is, it acts trivially in the internal degree of freedom subspace. The gaps between the bands are then small and the invariant-space decomposition is not useful. In a series of papers Bolte and Keppeler developed a semiclassical method relying on a flow where the qubit degree of freedom is represented on the sphere \cite{Bolte:2005ut, Bolte:1999vo, Bolte:1998ux}. A related method relying on a semiclassical flow in an extended phase space has been applied to cavity QED by \cite{Pletyukhov:2003vc, Pletyukhov:2002ux}. Here we use matrix flow in the standard Euclidean phase space to calculate the subleading order in the field dynamics which enables the analysis of the splitting of wave packets and generation of Schr\"odinger cat states. The qubit-field back action is studied in the extended phase-space framework in a forthcoming work \cite{glt}.

The method presented here is applicable to any initial state where the field is highly excited, not necessarily phase-space localized, including thermal states at temperature which are high compared with the mode frequency. The phenomena of interest in this work, namely the collapse of Rabi oscillations, wave packet splitting, and qubit-field entanglement, are captured by the $O(\varepsilon)$ equations, leading order qubit and subleading field. These effects and the related resonance phenomena take place at times of $O(\varepsilon^{-1/2})$, while the approximation scheme remains consistent for longer times $\ll\varepsilon^{-1}$. For times of $O(\varepsilon^{-1})$ however, the subleading terms become comparable with the leading terms, signaling the breakdown of the approximation. Indeed this is the time scale of revival of Rabi oscillation that are difficult to capture in semiclassical approximations based on phase-space flows \cite{Gat:2007de}. According to \cite{glt} the approximation based on the flow equations is valid at least for times $\ll\varepsilon^{-1/2}$, while the results of this paper suggest that this bound can be extended to $\ll\varepsilon^{-1}$, but further investigation of this issue is beyond the present scope.

The semiclassical framework is presented in section \ref{sec:fqi}, and is applied in section \ref{sec:res} to the case where the interaction is weak and resonant. The main calculational task here is the straightforward problem of the precession of the qubit polarization in a given periodic external field. This information is promoted using the semiclassical theory to a full description of the collapse of Rabi oscillations, the splitting of the initial field wave packet and the generation of qubit-field entanglement. There are natural resonances associated with each of these effects. The resonances are all degenerate in the RWA, and become distinct once nonresonant terms are taken into account. Our main result is the calculation of the resonant detuning for six independent quantum resonance criteria, using both weak interaction perturbation theory and numerical solution of the phase-space dynamical equations. The resonant detunings are shown in figure \ref{fig:shifts} as a function of interaction strength $\mu$, where the RWA results of zero detuning are recovered when $\mu=0$. The perturbative results are summarized in table \ref{tab:shifts}.

\section{Field-qubit interactions}\label{sec:fqi}
\subsection{Semiclassical Rabi dynamics}\label{sec:semi}
In the dipole approximation, the Hamiltonian of an electromagnetic field mode with an annihilation operator $a$ coupled to a qubit with a lowering operator $S$ takes the form
\begin{equation}\label{eq:rabi}
H=\hbar\omega \tfrac12(a^\dagger a+aa^\dagger)+\hbar\nu\tfrac12(S^\dagger S-SS^\dagger)+\hbar g(a+a^\dagger)(S+S^\dagger)
\end{equation}
where $a$ and $S$ satisfy the usual relations $[a,a^\dagger]=[S,S^\dagger]_+=1$, $[a,S]=S^2=0$, and $\omega$, $\hbar\nu$, and $\hbar g$ are the mode frequency, qubit energy splitting, and dipole coupling, respectively. The Hermitian components of the polarization vector operator $\Sv$ are defined by $\Sv_1=S+S^\dagger$, $\Sv_2=i(S-S^\dagger)$, and $\Sv_1\Sv_2=i\Sv_3$. We do not neglect the nonresonant terms $aS$ and $a^\dagger S^\dagger$ in the qubit-field interaction.

We assume that the initial state is a product wave packet $\ket{\alpha}\otimes\ket{\mathsf{p}}$, where $\ket{\alpha}$ is a phase-space localized wave packet, such as a coherent state, with $\bra{\alpha}a\ket{\alpha}=\alpha$, $|\alpha|\gg1$, and $\ket{\mathsf{p}}$ is an arbitrary pure qubit state, labeled by the direction of the polarization vector $\mathsf{p}$, so that $\Sv\cdot\mathsf{p}\ket{\mathsf{p}}=\ket{\mathsf{p}}$. It is convenient to define a small parameter $\varepsilon$ such that $\alpha=\bar\zeta/\sqrt\varepsilon$ with fixed $\bar\zeta$ and let $a=Z/\sqrt\varepsilon$ so that $[Z,Z^\dagger]=\varepsilon$ and
\begin{equation}
H=\tfrac{\hbar\omega}{\varepsilon} \tfrac12(Z^\dagger Z+ZZ^\dagger)+\hbar\nu\tfrac12(S^\dagger S-S^\dagger S)+\hbar \lambda(Z+Z^\dagger)(S+S^\dagger)
\end{equation}
Since the qubit-field interactions are typically weak, we let $\lambda=\frac{g}{\sqrt\varepsilon}$ be finite as $\varepsilon\to0$.

We can therefore write $H=\tfrac{1}{\varepsilon}\Hs+\Hv\cdot\Sv$, where $\Hs= \tfrac12\hbar\omega(Z^\dagger Z+ZZ^\dagger)$, $\Hv_1=\hbar \lambda(Z+Z^\dagger)$, $\Hv_2=0$, and $\Hv_3=\hbar\nu$, so that the Heisenberg equations are
\begin{align}
i\hbar\partial_t Z&=\tfrac{1}{\varepsilon}[Z,\Hs]+[Z,\Hv]\cdot\Sv\label{eq:dtZ}\\
\hbar\partial_t\Sv&=2\Hv\wedge\Sv\label{eq:dtS}
\end{align}

Our goal is solve the equations of motion order by order in $\varepsilon$. For this purpose we use the matrix-valued Wigner-Weyl representation \cite{Teufel:2003vf}, where an operator $B$ is represented by a a two-by-two matrix phase-space symbol $b(\zeta)$, where $\zeta$ is a complex coordinate in a phase-space with Poisson bracket $\{\zeta,\zeta^*\}=-i$, such that $Z$ and $\Sv$ are represented by 
\begin{equation} \label{eq:init}
z(\zeta)=\zeta\ ,\qquad\sv(\zeta)=\sigmav
\end{equation}
respectively, where $\sigmav$ is the vector of Pauli matrices. The product $B_1B_2$ of two operators is then represented by the Moyal product $b_1\#b_2=b_1b_2+\frac{i\varepsilon}{2}\{b_1,b_2\}+O(\varepsilon^2)$. The Weyl representation of the density matrix of the initial (pure) state is $\frac12(1+\sigmav\cdot\mathsf{p})W_\alpha(\zeta)$, where $W_\alpha(\zeta)$ is the Wigner function of the field state, so that quantum expectation values are given by
\begin{equation}\label{eq:expc}
\expc{B}=\int\frac{d^2\zeta}{2\pi\varepsilon}W(\zeta)\tr \Bigl(\frac{1+\sigmav\cdot\mathsf{p}}{2} b(\zeta)\Bigr)
\end{equation}
We assume that the squeezing of $\ket{\alpha}$ remains bounded as $\varepsilon\to0$, so that $W$ is localized near $\bar\zeta$ on the scale $\sqrt\varepsilon$.

The phase space representation of equations (\ref{eq:dtZ}--\ref{eq:dtS}) yields equations for the time dependent symbols $z_t$ and $\sv_t$ using the Moyal product rule
\begin{align}
\hbar\partial_t z&=\{z,\hs\}+\varepsilon\{z,\hv\}\cdot\sv\label{eq:dtz}\\
\hbar\partial_t\sv&=2\hv\wedge\sv\label{eq:dtSs}
\end{align}
keeping only the leading term in equation (\ref{eq:dtSs}). Assuming small $\varepsilon$ asymptotics
\begin{equation}\label{eq:asym}
z_t=z_t^{(0)}+\varepsilon z_t^{(1)}+\cdots\ ,\qquad\sv_t=\sv_t^{(0)}+\cdots
\end{equation}
we obtain
\begin{align}
\hbar\partial_t z_t^{(0)}&=-i\partial_{\zeta^*}\hs^{(0)}\label{eq:z0}\\
\hbar\partial_t\sv_t^{(0)}&=2 \hv^{(0)}\wedge\sv_t^{(0)}\label{eq:s0}\\
\hbar\partial_t z_t^{(1)}&=-i\partial_{\zeta^*}\hs^{(1)}-i\partial_{\zeta^*}\hv^{(0)}\cdot\sv^{(0)}\label{eq:z1}
\end{align}
where the asymptotic expansions of the phase-space functions on the right-hand sides are defined by 
\begin{displaymath}
f(z_t^{(0)}+\varepsilon z_t^{(1)}+\cdots)=f^{(0)}(\zeta)+\varepsilon f^{(1)}(\zeta)+\cdots
\end{displaymath}
 for $f=\hv$, $\partial_{\zeta^*}\hs$, and $\partial_{\zeta^*}\hv$. Here $\hs$ and $\hv$ are the symbols of $\Hs$ and $\Hv$, respectively.  

The asymptotic approximation (\ref{eq:asym}) is not uniform in time. Nevertheless, the solution of equations (\ref{eq:z0}--\ref{eq:z1}) derived below is consistent with the approximation for all times $\ll\varepsilon^{-1}$, and captures all the  known phenomena of wave packet cavity QED in this time interval. We therefore proceed under the conjecture that the validity of the approximation can be extended to all orders for $t\ll\varepsilon^{-1}$.

Note that unlike the semiclassical equations of motion generated by many-component Hamiltonians whose principal symbol is non-scalar \cite{Panati:2003ul,Stiepan:2012kl}, equation (\ref{eq:z0}--\ref{eq:z1}) are not flow equations of a classical Hamiltonian. In a companion work \cite{glt} we derive Hamiltonian equations of motion in an extended phase space that approximate the Heisenberg picture dynamics for times of $o(\varepsilon^{-1/2})$.

\subsection{Rabi oscillations and field-qubit entanglement}
The results of the previous subsection mean that CQED expectation values for highly excited initial states may be calculated using (\ref{eq:expc}) by solving equations (\ref{eq:z0}--\ref{eq:z1}).

Equation (\ref{eq:z0}) is trivially solved by $z_t(\zeta)=\zeta e^{-i\omega t}$, so that $\hs^{(0)}=\hbar\omega|\zeta|^2$ and
\begin{equation}
 (\hv^{(0)}_1,\hv^{(0)}_2,\hv^{(0)}_3)=(\hbar\mu\cos(\omega t-\varphi),0,\tfrac12\hbar\nu)
\end{equation}
 where $\mu=\lambda|\zeta|$ and $\varphi=\arg\zeta$.
Substituting in (\ref{eq:s0}) gives a linear homogeneous system for $\sv$ with periodic coefficients. It is identical with the well-studied equations for the Rabi oscillations of a two-level system in a classical radiation field \cite{Bloch:1940wy,Rabi:1939tl}.
Here it is convenient to apply Floquet theory \cite{Autler:1955uf} that implies that the system has three quasi-periodic fundamental solutions 
\begin{equation}\label{eq:fund}
e^{i\Omega_kt}\sper_k(t)\ ,\qquad k=-1,\,0,\,1\ ,\qquad \sper_k(t+\tfrac{2\pi}{\omega})=\sper_k(t)
\end{equation}
The Floquet frequencies are defined modulo $\omega$; for concreteness, we choose the branch $|\Omega_k|\le\frac\omega2$. Because the RHS of (\ref{eq:s0}) is a cross product, the Rabi frequency $\Omega\equiv\Omega_1=-\Omega_{-1}$ is real, $\Omega_0=0$, and $\sper_{-k}(t)=\sper_k(t)^*$ (complex conjugate).

The solution of (\ref{eq:s0}) can be expressed in the standard way as a linear combination of the fundamental solutions (\ref{eq:fund}): Using subscripts $a$ and $b$ for the polarization index, the fundamental solutions form a 3-by-3 matrix $\sper_{ak}(t)$; denoting the components of the inverse of $\sper(t=0)$ by $\sper^{-1}_{kb}$ we have
\begin{equation}\label{eq:vv}
\sv_{t,a}=\sum_{kb}e^{i\Omega_k t}\sper_{ak}(t)\sper^{-1}_{kb}\sigma_b\equiv\sum_{b}\mathsf{O}_{ab}(t)\sigma_b
\end{equation}
defining the orthogonal matrix $\mathsf{O}$, which describes the Rabi oscillations of the polarization vector of a qubit in a classical radiation field. The Rabi oscillations are quasi-periodic with base frequencies $\Omega$ and $\omega$; $\Omega$-periodic oscillations are obtained only in the RWA.

Using (\ref{eq:expc}) we now obtain  
\begin{equation}
\expc{\sv_a}_t=\sum_{kb}\int\frac{d^2\zeta}{2\pi\varepsilon}W_\alpha(\zeta)e^{i\Omega_k t}\sper_{ak}(t)\sper^{-1}_{kb}\mathsf{p}_b
\end{equation}
Recall that $\sper(t)$ and $\Omega$ depend parametrically on $\zeta$ through $\mu$ and $\varphi$, and this determines the $\zeta$ dependence of $\sv_t$.
The symbol $\sv_t(\zeta)$ therefore varies slowly on the domain of size $\sqrt\varepsilon$ where $W_\alpha$ is appreciable, and we may write
\begin{equation}
\Omega(\zeta)\sim\Omega(|\bar \zeta|)+(|\zeta|-|\bar \zeta|)\Omega'(|\bar \zeta|)\equiv\bar\Omega+(|\zeta|-|\bar \zeta|)\bar\Omega' \ ,\quad \sper_{ak}(t;\zeta)\sim \sper_{ak}(t,\bar \zeta)\equiv\bar \sper_{ak}
\end{equation}
so that
\begin{equation}
\expc{\sv_a}_t=\sum_{kb}\bar \sper_{ak}(t)\bar \sper_{kb}^{-1}\mathsf{p}_be^{i\bar\Omega_k t}\int\frac{d^2\zeta}{2\pi\varepsilon}W_\alpha(\zeta)e^{i(|\zeta|-|\bar \zeta|)\bar\Omega_k't}
\end{equation}
or, defining the normalized real-variable function  $w(|\zeta|-|\bar \zeta|)=\int \frac{|\zeta|d\arg \zeta}{2\pi\varepsilon}W_\alpha(\zeta)$,
\begin{equation}\label{eq:tc}
\expc{\sv_a}_t=\sum_{kb}\bar \sper_{ak}(t)\bar \sper_{kb}^{-1}\mathsf{p}_be^{i\bar\Omega_k t}\int dx w(x)e^{i\bar\Omega_k'tx}
\end{equation}
The integral on the RHS of (\ref{eq:tc}) is the Fourier transform of $w$. The function $w(x)$ is localized near zero on the scale $\sqrt{\varepsilon}$, so its Fourier transform decays on scale $t\sim\varepsilon^{-1/2}$. 
Therefore for short times $t\ll\varepsilon^{-1/2}$ the subleading $\bar\Omega_k'$ is negligible and 
\begin{equation}
\expc{\sv}_t=\bar{\mathsf{O}}(t)\mathsf{p}
\end{equation}
That is, the polarization of the wave packet Rabi oscillates as a whole with a frequency determined by its center.

For long times $t\gg\varepsilon^{-1/2}$ on the other hand, $\int dx w(x)e^{i\bar\Omega_k'tx}\ll1$ so that  
\begin{equation}\label{eq:stl}
\expc{\sv}_t=\bar{\mathsf{Q}}(t)\mathsf{p}\ ,\qquad \bar{\mathsf{Q}}_{ab}(t)=\bar \sper_{a0}(t)\bar \sper^{-1}_{0b}
\end{equation}
and the Rabi frequency $\bar\Omega$ disappears from the polarization dynamics. This is the generalization of the collapse of Rabi oscillations \cite{Eberly:1980tj}; unlike the full collapse in the RWA, in general the polarization continues to oscillate periodically with frequency $\omega$ in the long-time regime.

The wavepacket polarization state is an indicator of the entanglement between the qubit and the field. Since the field-qubit state is pure, the purity $\tr\rho_\text{qubit}^2$ of the field-reduced density matrix $\rho_\text{qubit}=\tr_\text{field}\rho$ is an entanglement monotone \cite{Peres:1995ti}, with $\tr\rho_\text{qubit}^2=\frac12$ when the field and qubit are maximally entangled, and $\tr\rho_\text{qubit}^2=1$ when the state is separable. Since $\rho_\text{qubit}=\frac12(1+\expc{\sv}\cdot\sigmav)$
\begin{equation}\label{eq:ent}
\tr\rho_\text{qubit}^2=\tfrac12(1+\expc{\sv}^2)
\end{equation}
so that $\expc{\sv}^2$ is an entanglement monotone as well. In the short time regime $t\ll\varepsilon^{-1/2}$ the wave packet is almost fully polarized 
since $\mathsf{O}$ is orthogonal, and the state remains almost separable. In contrast, $\mathsf{Q}$ is not orthogonal and therefore the field and qubit are in general entangled and the degree of entanglement oscillates with frequency $\omega$. The details of the entanglement dynamics are studied below.

\subsection{Wave packet splitting}
The analysis presented so far was based on the leading order equations of motion (\ref{eq:z0}--\ref{eq:s0}) that  hold for classical electromagnetic radiation, and the wave packet effects were derived by combining them with phase-space integration. We turn next to the subleading field flow equation (\ref{eq:z1}). Since $\partial_{\zeta^*}\hs=\zeta$, and $\partial_{\zeta^*}\hv_1=\lambda$, $\partial_{\zeta^*}\hv_{2,3}=0$, we have 
\begin{equation}
\partial_t z_t^{(1)}=-i\omega z_t^{(1)}-2i\lambda \sv_1
\end{equation}
so that $z_t^{(1)}$ has non-trivial qubit action.

We are interested in the long time $t\gg\varepsilon^{-1/2}$ behavior of $z_t^{(1)}$, so we use the late-time solution (\ref{eq:stl}) for $\sv$ and Fourier expand $\sper$
\begin{equation}
\sper_{ak}(t)=\sum_{n=-\infty}^\infty\tsper_{k,na}e^{-in\omega t_\varphi}\ ,\qquad t_\varphi= t-\frac{\varphi}{\omega}
\end{equation}
to obtain
\begin{align}\label{eq:ztld}
z_t^{(1)}&=-2i\lambda e^{-i\omega t}\int_0^tdt'\sv_1(t')e^{i\omega t'}\nonumber\\&=
-2i\lambda e^{-i\omega t}\sum_b\Bigl(\tsper_{0,1}\sper^{-1}_{0b}t+\sum_{(k,n)\ne(0,1)}\frac{\tsper_{k,n1}\sper^{-1}_{kb}e^{ik\Omega t-i(n-1)\omega t_\varphi}}{i(k\Omega+(n-1)\omega)}\Bigr)\sigma_b
\end{align}
All the terms in parenthesis in this expression except the first are bounded. For late times, 
therefore, the first term is dominant and we can write
\begin{equation}\label{eq:vt}
z_t=e^{-i\omega t}(\zeta+ vt\mathsf{n}\cdot\sigmav)
\end{equation}
up to higher order terms, with 
\begin{equation}\label{eq:splitz}
v=-2i\varepsilon\lambda e^{i\varphi}\tsper_{0,1}\|\sper_0^{-1}\|\,,\ \|\sper_0^{-1}\|^2=\sum_b(\sper^{-1}_{0b})^2\,,\ \text{and}\ \mathsf{n}_b=\sper^{-1}_{0b}/\|\sper_0^{-1}\|
\end{equation}

As an operator in the qubit space $z_t$ has eigenvalues $e^{-i\omega t}(\zeta\pm vt)$ with corresponding eigenprojections $\frac12(1\pm\mathsf{n}\cdot\sigmav)$. It follows that the expectation value of a classically smooth field observable $B$ with symbol $b(\zeta)$ is 
\begin{equation}\label{eq:splitw}
\expc{B}=\int \frac{d^2\zeta}{2\pi\varepsilon}\Bigl(b(e^{-i\omega t}(\zeta+vt))\frac{1+\mathsf{p}\cdot\mathsf{n}}{2}+b(e^{-i\omega t}(\zeta-vt))\frac{1-\mathsf{p}\cdot\mathsf{n}}{2}\Bigr)W(\zeta)
\end{equation}
which means that the initial wave packet splits into two fragments with statistical weights $\frac12(1\pm\mathsf{p}\cdot\mathsf{n})$ that drift with speeds $\pm vt$ in a frame rotating with frequency $\omega$. 

The splitting is a necessary condition for appreciable field-qubit entanglement; here splitting and entanglement arise on the same time scale $t_c\sim\varepsilon^{-1/2}$. The solution (\ref{eq:vt}) is valid for time $t\gg t_c$ and then $vt$ is much larger than the field uncertainty in the initial wave packet and the field state consists of two well-separated sub wave packets. For even later times $t\sim\varepsilon^{-1}$ the subleading term in (\ref{eq:vt}) becomes comparable with the leading term signaling a breakdown of the approximation at this time scale. Our analysis is therefore limited to the range $t\ll\varepsilon^{-1}$.

\subsection{Calculation of the polarization dynamics}
The polarization dynamics Floquet problem is equivalent to an infinite matrix eigenproblem for the Floquet function Fourier coefficients \cite{Autler:1955uf}
\begin{equation}\label{eq:flo}
\sum_{mb} L_{na,mb}\tsper_{k,mb}=i\Omega_k\tsper_{k,na}
\end{equation}
where the nonzero components of the matrix $L$ are $L_{na,na}=i\omega n$, $L_{n2,n1}=-L_{n1,n2}=\nu$, and $L_{n3,(n+1)2}=L_{(n+1)3,2}=-L_{n2,(n+1)3}=-L_{(n+1)2,3}=\mu$. We present results based on perturbative and numerical solutions of the eigenproblem.

When $\mu=0$ the matrix $L$ is block-diagonal. For small $\mu$ the Floquet functions and Rabi frequency can therefore be calculated by perturbation theory. As our main goal is to study resonance phenomena we assume that the qubit-field detuning $\delta=\nu-\omega$ is small, and then the spectrum has gaps of order $\delta$. We therefore assume that $\delta=O(\mu)$ and shift a block-diagonal term proportional to $\delta$ to the perturbation matrix, so that the unperturbed spectrum consists of all integer multiples of $i\omega$, each 3-fold degenerate.  
We calculated the Floquet functions and Rabi frequency to third order in $\mu$ using the degenerate perturbation theory formulation of \cite{Suzuki:1983uy}. 
The leading order $O(\mu)$ perturbation that includes the interaction between the $n=0$ and the $n=\pm1$ blocks is equivalent to the RWA. Higher orders go beyond the RWA approximation, and are found to be numerically satisfactory  up to at least $\mu=\frac12$.

In addition, we solved equation (\ref{eq:flo}) numerically by truncating the blocks of $L$ with $|n|>n_\text{max}$. Very precise solutions were obtained for $\mu\le\frac12$ with $n_\text{max}=8$ or smaller, where it is easy to diagonalize $L$ numerically. Higher values of $\mu$ can also be studied in this method, but since resonance phenomena are less important when $\mu$ is large we did not pursue this direction.
\renewcommand{\arraystretch}{1.5}
\begin{figure}
\includegraphics[width=15cm]{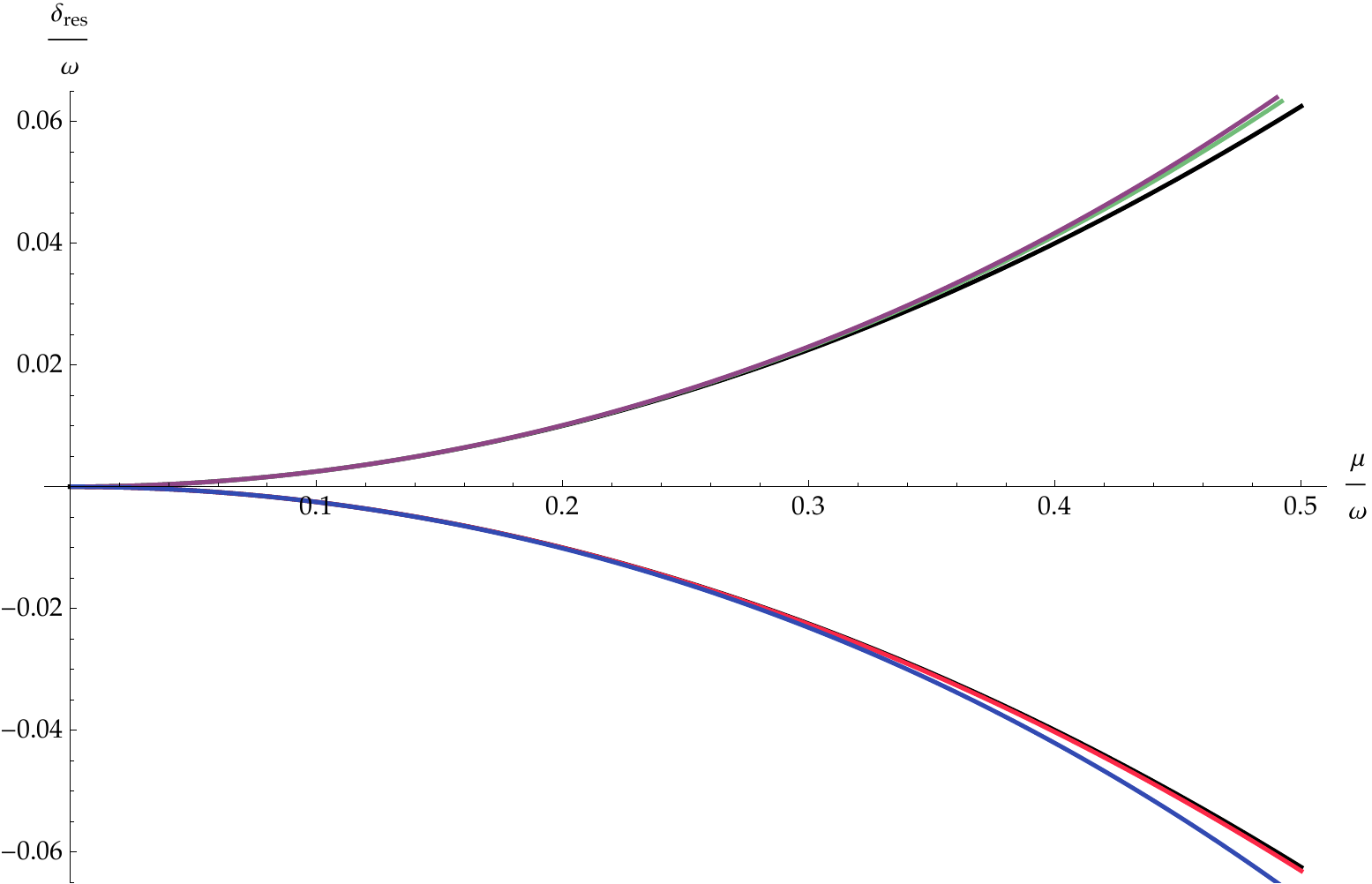}
\caption{The resonant detuning $\delta_\text{res}$ as a function of the interaction strength $\mu$, both expressed in units of the field mode frequency $\omega$. The classical Bloch-Siegert detuning $\delta_\text{BS}$ is shown in blue, and the quantum detunings $\delta_\text{tc}$, $\delta_\text{fc}$, $\delta_\text{rc}$, $\delta_\text{en}$, $\delta_\text{vs}$, and $\delta_\text{ws}$, are shown in green, blue, violet, red, green, and violet, respectively. $\delta_\text{fc}$, $\delta_\text{vs}$, and $\delta_\text{ws}$, are not graphically distinguishable from $\delta_\text{BS}$, $\delta_\text{tc}$, and $\delta_\text{rc}$ (respectively), but they are all numerically distinct. The graphs of the leading term in the expansion of the resonant detunings $\delta_\text{res}=\pm\frac14\mu^2$ are shown in black. The various resonant detunings are defined in section \ref{sec:res}. \label{fig:shifts}
}
\end{figure}

\section{Resonance phenomena}\label{sec:res}
In many experiments the field-qubit interaction energy is much smaller than both the qubit energy splitting and the electromagnetic mode energy quantum. In these cases the only way to achieve appreciable energy exchange between the field and the qubit is to make the interaction \emph{resonant}, so that the  detuning $\delta=\nu-\omega$ is comparable with or smaller than the interaction frequency scale $\mu$. 

\begin{figure}
\includegraphics[width=16cm]{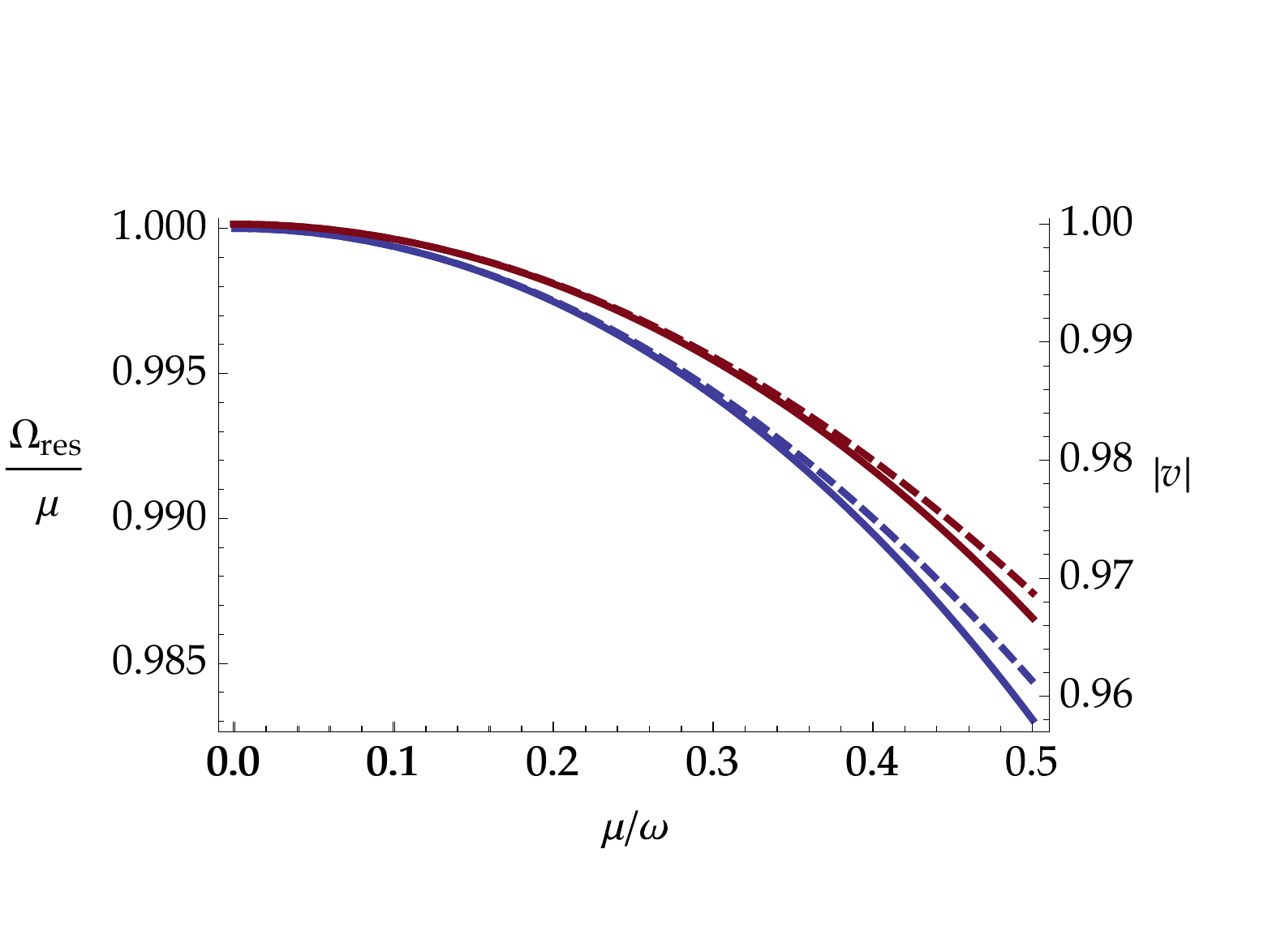}
\caption{Blue tinted curves: The resonant Rabi frequency $\Omega_\text{res}$ in units of its value $\Omega_\text{RWA}=\mu$ in the rotating wave approximation. Red tinted curves: The magnitude of the drift speed $|v|$ of the wave packet fragments in units of $\frac{\varepsilon\mu}{|\bar\zeta|}$. Both are shown as a function of the normalized interaction strength $\frac{\mu}{\omega}$. The perturbative and the numerical results are shown in dashed and solid lines, respectively.\label{fig:bs-sp}
}
\end{figure}

When $\mu/\omega\to0$, the rotating wave approximation becomes exact, and the resonance condition $\delta=0$ implies that the visibility of the Rabi oscillations is maximal, and the Rabi frequency $\Omega$ is minimal for a fixed $\mu$. When $\mu/\omega$ is small but finite the resonance shifts to nonzero detuning 
\begin{equation}
\frac{\delta_\text{res}}{\omega}=c\Bigl(\frac{\mu}{\omega}\Bigr)^2+\cdots\end{equation}
Bloch and Siegert \cite{Bloch:1940wy} calculated perturbatively the shift of the minimum of $\Omega$ as a function of $\frac{\delta}{\omega}$ for a given $\frac{\mu}{\omega}$ and showed that $c=-\frac14$ and $\Omega_\text{res}=\mu(1-\frac{1}{16}(\frac{\mu}{\omega})^2+\cdots)$. The Bloch-Siegert problem has been revisited several times and high-order terms have been calculated \cite{Swain:1973wy,CohenTannoudji:1973wya}. It is straightforward to calculate the Bloch-Siegert shift from a numerical solution of the Floquet equation (\ref{eq:flo}); the perturbative and numerical values of the Bloch-Siegert detuning are shown and compared in figure \ref{fig:shifts} and those of $\Omega_\text{res}$ are shown in figure \ref{fig:bs-sp}.
\begin{table}
\centering{\begin{tabular}{|l|c|c|c|c|c|c|c|}
\hline
Resonant detuning&$\delta_\text{BS}$ & $\delta_\text{tc}$ & $\delta_\text{fc}$ & $\delta_\text{rc}$ & $\delta_\text{en}$ & $\delta_\text{vs}$ & $\delta_\text{ws}$\\
\hline
Shift coefficient $c$ & $-\frac14$ & $\frac14$ & $\frac14$ & $-\frac14$ & $-\frac14$ &  $\frac14$ & $-\frac14$\\
\hline
\end{tabular}}
\caption{Coefficient of the leading order detuning $\delta_\text{res}=c\mu^2$ for small $\mu$, of the various wave packet-qubit resonances. $c=\pm\frac14$ in all cases. The resonances are defined in section \ref{sec:res}.\label{tab:shifts}}
\end{table}

The Rabi frequency is defined by the qubit response to external fields, and the Bloch-Siegert resonance therefore characterizes the classical limit $\varepsilon\to0$. We show that when $\varepsilon$ is nonzero, several additional resonances 
arise from the processes tied to the quantum properties of the electromagnetic field.
In the rotating wave approximation all the resonances occur for zero detuning. In the remainder of this section we study these resonances and calculate the resonant detuning of each type. Interestingly, we find that while the resonances are all distinct, the leading order detuning of the seven resonances (including the Bloch-Siegert resonance) fall under the two cases $c=\pm\frac14$. The values of the  resonant detunings are summarized in table \ref{tab:shifts} and figure \ref{fig:shifts}.

\subsection{Collapse of Rabi oscillations}
\begin{figure}
\includegraphics[width=16cm]{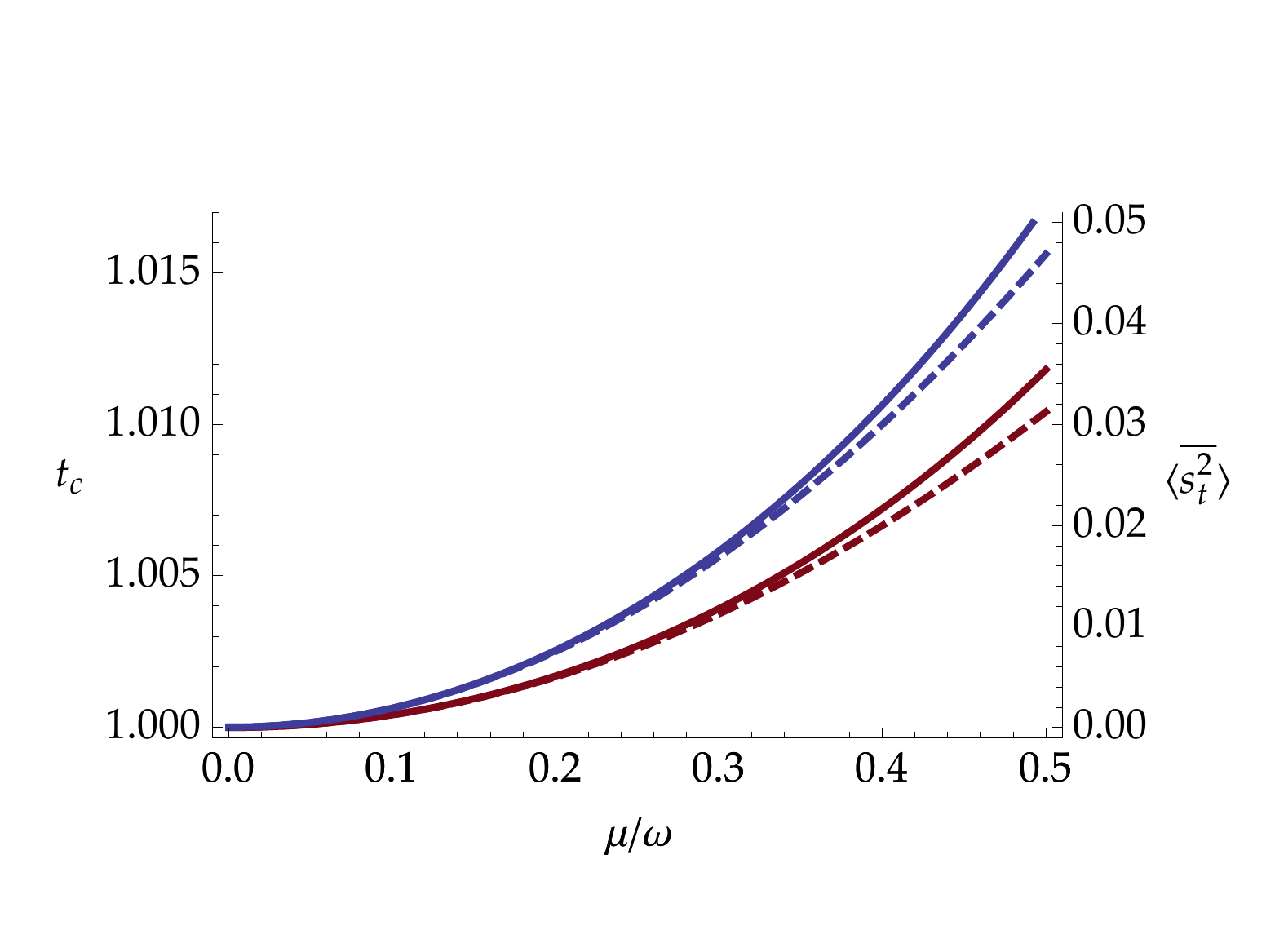}
\caption{Blue tinted curves: The resonant collapse time $t_c$ in units of the RWA collapse time. Red tinted curves: The expectation value of the square of the qubit polarization vector $\sv$, a decreasing entanglement monotone, averaged over a period $2\pi/\omega$ at times much later than the collapse time $t_c$. Both are shown as a function of the normalized interaction strength $\frac{\mu}{\omega}$. The perturbative and numerical results are shown in dashed and solid lines, respectively.\label{fig:ent-tc}
}
\end{figure}

It follows from equation (\ref{eq:tc}) that the time scale $t_c$ of the collapse of Rabi oscillations is inversely proportional to $\partial_{|\zeta|}\Omega=\frac{\mu}{|\bar\zeta|}\partial_\mu \Omega$. In the RWA $(\partial_\mu\Omega)^{-1}=\frac{\sqrt{\mu^2+\delta^2}}{\mu}$ is minimal at resonance $\delta=0$, while for non-zero $\mu$ the minimal collapse time, obtained for $\delta_\text{tc}$ with $c=\frac14$, is
\begin{equation}
t_c=t_{c,\text{RWA}}\Bigl(1+\frac{1}{16}\Bigl(\frac{\mu}{\omega}\Bigr)^2+\cdots\Bigr)
\end{equation}
The numerically calculated $\delta_\text{tc}$ and $t_c$ are shown in figures \ref{fig:shifts} and \ref{fig:ent-tc}, respectively.

If the initial qubit is a population eigenstate, so that $\mathsf{p}=\mathsf{e}_3$ is the direction 3 unit vector, the collapsed population difference in the RWA is
$\expc{\sv_{3,t}}=\frac{\delta^2}{\mu^2+\delta^2}$ so that
exact resonance also implies complete collapse of the population. When $\mu\ne0$ the Rabi oscillations \emph{fully} collapse with $\expc{\sv_{3,t}}=0$ identically only for resonant detuning $\delta_\text{fc}$ with $c=\frac14$. Otherwise the polarization continues to oscillate with frequency $\omega$ after the collapse of Rabi oscillations; still, for detuning $\delta_\text{rc}$ with $c=-\frac14$ the  polarization oscillates with zero mean, $\overline{\expc{\sv_{3,t}}}\equiv\int_{t_0}^{t_0+2\pi/\omega}dt\expc{\sv_{3,t}}=0$. The numerically calculated $\delta_\text{fc}$ and $\delta_\text{rc}$ are shown in figure \ref{fig:shifts}.

\subsection{Field-qubit entanglement}
The resonant Rabi dynamics is  characterized in the RWA by maximal entanglement of the field and atom, so that $\expc{\sv}=0$ after the collapse of the Rabi oscillations. When $\mu$ is nonzero, $\expc{\sv}$ does not tend to zero identically for any choice of $\delta$ and continues to oscillate with frequency $\omega$. We therefore define the entanglement-based resonance where the mean square polarization $\overline{\expc{\sv_{t}}^2}\equiv\int_{t_0}^{t_0+2\pi/\omega}dt\expc{\sv_{t}}^2$ is minimal (see equation (\ref{eq:ent})). It is obtained for $\delta_\text{en}$ with $c=-\frac14$ where
\begin{equation}
\overline{\expc{\sv_{t}}^2}=\frac{\mu^2}{8}+\cdots
\end{equation}
The numerically calculated detuning and mean square polarization are shown in figures \ref{fig:shifts} and \ref{fig:ent-tc}, respectively.

\subsection{Splitting of a localized wave packet}
The speed $|v|$ of the split wave packet fragments is proportional to $\tsper_{0,1}\|\sper_0^{-1}\|$ according equation (\ref{eq:splitz}). In the RWA $\tsper_{0,1}\|\sper_0^{-1}\|=\frac{\mu}{2\sqrt{\mu^2+\delta^2}}$
so that $|v|$ is maximal on resonance. The maximum speed is shifted to $\delta_\text{vs}$ with $c=\frac14$ for nonzero $\mu$, where  
\begin{equation}
|v|=\frac{\varepsilon\mu}{|\bar\zeta|}\Bigl(1-\frac{\mu^2}{8}+\cdots\Bigr)
\end{equation}
The numerically calculated $\delta_\text{vs}$ and $|v|$ are shown in figures \ref{fig:shifts} and \ref{fig:bs-sp}, respectively.

For an initial population eigenstate $\mathsf{p}=\mathsf{e}_3$, it follows from equation (\ref{eq:splitw}) that the wave packet is split into fragments of equal statistical weight when $\mathsf{n}_3=0$. In the rotating wave approximation $\mathsf{n}_3=\frac{\mu\delta}{\mu^2+\delta^2}$ so that the resonance condition implies equal weight splitting of the initial wave packet. Equal weight splitting occurs at $\delta_{\text{ws}}$ with $c=\frac14$ for nonzero $\mu$. The  numerically calculated $\delta_\text{ws}$ is shown in figure \ref{fig:shifts}.

\section{Conclusions}
The quantum dynamics of the full Rabi dynamics of a qubit coupled to an electromagnetic mode is considerably more complicated than the simplified dynamics in the rotating wave approximation, and accordingly less understood. For highly-excited wave packets involving a large number of eigenstates, the semiclassical method presented here yields accurate quantum expectation values for long times, where the dynamics departs radically from that of a qubit in a classical radiation field. It therefore enables us to study the collapse of Rabi oscillations, field-qubit entanglement and wave packet splitting effects that characterize the interaction of a qubit and a highly-excited wave packet on the basis of phase-space dynamics of two-by-two matrices that operate in the qubit Hilbert space. An alternative semiclassical method based on a scalar flow in an extended phase space is developed in \cite{glt}.

The phase-space equations of the full Rabi dynamics are themselves non-trivial. Still, they are readily amenable to perturbative analysis near integrable limits and easy to solve numerically. The solution suffices for the computation of the quantum observables of interest. The focus here has been on the resonant limit, where the interaction between the field and qubit is weak and the detuning between the field mode and the qubit frequencies is small; the rotating wave approximation becomes exact in the limit that both quantities tend to zero together.

Bloch and Siegert observed \cite{Bloch:1940wy} that for a finite-strength interaction with classical radiation the center of the resonance is shifted to a finite detuning value. In quantum electrodynamics, the picture changes in two ways: first of all, several concepts of resonance arise, each associated to a certain characteristic observable; secondly, unlike the Bloch-Siegert shift, some of the resonances shift to \emph{positive} detuning.


We find that the center shift of all resonances take only two possible values in the leading order (in qubit-field interaction strength), which are the negatives of each other, as summarized in table \ref{tab:shifts} and shown in figure \ref{fig:shifts}. This partial lifting of the degeneracy is difficult to understand purely on the basis of the phase-space dynamics, but seems more natural in view of the approximate transformation \cite{Beaudoin:2011ip} of the Rabi Hamiltonian (\ref{eq:rabi}) to RWA form that depends on a single additional parameter.

Nevertheless, the resonances are all distinct, and split apart once higher order terms in the interaction are taken into account, although the splitting of some of the shifts is so small that it cannot be observed on the scale of the plots shown here. The comparison between the perturbative and numerical calculations shows that although the perturbation theory is ostensibly valid only for weak interactions, the leading order in the perturbation series is a numerically excellent approximation for all interaction values up to half the field mode frequency, at which point the notion of resonance is no longer sharp.

The main limitation of the present theory is that it is not uniform in time, a  property common to all dynamical semiclassical approximations. We showed that it is possible to extend it consistently for times long enough to cover several non-trivial quantum electrodynamic effects and analyze the associated resonance phenomena. However, some of the wave packet phenomena cannot be captured, notably the revival of Rabi oscillations. Such long time effects may be studied using wave packet semiclassical methods like those employed in \cite{Leshem:2011ji}. The advantage of phase-space-flow-based theories like the present one is that they apply to any highly-excited initial state and in particular are not limited to localized wave packets.

The experimental observation of the present results is not straightforward, since it entails very strong qubit-field interactions as well as long time coherence of excited wave packets. On the other hand, the quantum effects arise faster with strong interactions, so the necessary coherence time is shorter. A clear indication of a quantum electrodynamic resonance would then be a blue shifted resonance curve.
Continuous-variable quantum information processing in superconducting quantum circuits is a possible application, where resonance conditions may arise when seeking resource optimization.

This work was supported by the German-Israeli Foundation (grant 980-184.14/2007).

\bibliographystyle{unsrt}
\bibliography{/Users/omrigat/research/library}

%
%
%
%
%
%
%

\end{document}